\documentclass[aps,pra,twocolumn,superscriptaddress,notitlepage,nofootinbib,longbibliography, colorlinks=true]{revtex4-2}
\usepackage{amsmath,amssymb,amsfonts,graphicx,float,times,psfrag}
\usepackage[pdftex]{color}
\usepackage{amsmath,bm}
\usepackage[colorlinks, linkcolor=blue, citecolor=blue,  breaklinks=true]{hyperref}
\usepackage{amssymb,amsmath,graphicx,bm,float}  
\usepackage{mathtools,amsfonts,mathptmx}
\usepackage[utf8]{inputenc}
\usepackage[T1]{fontenc}
\usepackage{xcolor}
\usepackage{braket}
\usepackage{siunitx}
\usepackage{booktabs}
\usepackage[titletoc,title]{appendix}

\usepackage[shortlabels]{enumitem}

\usepackage[capitalize,noabbrev]{cleveref}
\Crefname{equation}{Eq.}{Eqs.}
\crefname{equation}{Equation}{Equations}
\Crefname{figure}{Fig.}{Figs.}
\crefname{figure}{Figure}{Figures}
\begin{document}
	\title{Quantum phase synchronisation enhanced via Coulomb interaction in an optomechanical system }
	
	\author{E. K. Berinyuy}
	\email{emale.kongkui@facsciences-uy1.cm}
	\affiliation{Department of Physics, Faculty of Science, University of Yaounde I, P.O.Box 812, Yaounde, Cameroon}
	
		\author{P. Djorwé}
	\email{djorwepp@gmail.com}
	\affiliation{Department of Physics, Faculty of Science, 
		University of Ngaoundere, P.O. Box 454, Ngaoundere, Cameroon}
		
	\author{J.-X. Peng}
	\affiliation{School of Physics and Technology, Nantong University, Nantong, 226019, People’s Republic of China}
	
	\author{A. Sohail}
	\affiliation{Department of Physics, Government College University, Allama Iqbal Road, Faisalabad 38000, Pakistan}
	
	\author{J. Ghosh}
	\affiliation{School of Nanoscience and Technology, IIT Kharagpur, West Bengal 721302, India}
	
	\author{A.-H. Abdel-Aty}
\email{amabdelaty@ub.edu.sa}
\affiliation{Department of Physics, College of Sciences, University of Bisha, Bisha 61922, Saudi Arabia}
	
	\author{S. G. N. Engo}
	\email{serge.nana-engo@facsciences-uy1.cm}
	\affiliation{Department of Physics, Faculty of Science, University of Yaounde I, P.O.Box 812, Yaounde, Cameroon}
	
\author{S. K. Singh}
	\affiliation{Department of Physics, Akal University, Talwandi Sabo, Bathinda 151302, India}
\begin{abstract}
	In this work, we investigate the dynamics of quantum synchronization in a four-mode optomechanical system, focusing on the influence of the Coulomb interaction between two mechanical resonators. We analyze the effect of the Coulomb coupling on three distinct synchronization regimes, i.e., complete quantum synchronization, $\phi$-synchronization, and quantum phase synchronization. Our results show that while the Coulomb interaction plays a pivotal role in significantly enhancing quantum phase synchronization by facilitating energy exchange and phase coherence, it has little impact on complete and $\phi$-synchronization. This indicates that amplitude and frequency locking are primarily determined by the optical driving, whereas phase alignment depends critically on inter-resonator coupling. We also demonstrate that the oscillations of the two optical cavities, which are indirectly coupled via the mechanical resonators, can become aligned over time, resulting in classical synchronization. These findings provide a robust mechanism for controlling collective quantum dynamics and offer a foundation for applications in quantum communication, precision sensing, and the development of synchronized quantum networks.
\end{abstract}

	\maketitle

\section{Introduction} \label{sec:Intro}
Optomechanical systems, which explore the interaction between electromagnetic fields and mechanical motion via radiation pressure, have emerged as a thriving platform for studying quantum phenomena at the macro- and mesoscopic scales~\cite{Aspelmeyer2014,Shlomi2015}. These systems typically consist of optical cavities coupled to mechanical resonators, enabling the transfer of energy and information between light and matter~\cite{EMALE2025,kongkui2025}. The inherent nonlinearity of the radiation-pressure interaction allows rich dynamics, including cooling of mechanical motion to the quantum ground state~\cite{Chan2011,teufel2011}, quantum state transfer~\cite{wang2012}, and the generation of non-classical states~\cite{riedinger2016,Ema2025,Massem2025}. Over the past decade, experimental advances have made it possible to observe quantum effects in optomechanical devices, opening new pathways toward quantum technologies such as ultrasensitive sensing~\cite{Djor2019,fogliano2021,Djor2024,Tang2025}, quantum memory~\cite{shandilya2021,sete2015,wallucks2020}, and hybrid quantum networks~\cite{dong2015,Moli2022}.
Within the broader context of quantum dynamics, synchronization that is a fundamental phenomenon where coupled systems adjust their rhythms due to weak interaction has attracted significant theoretical and experimental interest. While synchronization is well-understood in classical systems \cite{Djor2020,Li2022,Rodrigues2021}, its quantum counterpart presents unique challenges and opportunities due to coherence, entanglement, and quantum noise. Quantum synchronization has been studied in various platforms, including trapped ions, spin systems, and optomechanical arrays~\cite{PhysRevA.109.023502,PhysRevA.99.033818,PhysRevE.95.022204,doi:10.1080/09500340.2016.1249977,Peng2025,Joy}. In particular, the synchronization of mechanical oscillators or optical modes via shared fields or direct coupling offers a promising route to achieve correlated quantum dynamics, with potential applications in quantum communication~\cite{PhysRevE.95.022204} and synchronous quantum information processing.

Despite progress, most studies on quantum synchronization in optomechanical systems have focused on linearized interactions mediated by photons or phonons~\cite{Sun2024}. Less attention has been paid to the role of direct mechanical interactions, such as Coulomb forces, which can induce strong and tunable coupling between mechanical elements. Such interactions not only enrich the system dynamics but also provide a new mechanism for achieving and controlling synchronization without relying solely on optical mediation. This motivates the exploration of hybrid optomechanical setups where charged mechanical resonators interact via Coulomb forces, thereby enabling synchronization through both photonic and electrostatic channels.

In this work, we investigate quantum synchronization in a hybrid optomechanical system comprising two optical cavities, each coupled to a mechanical oscillator, with the mechanical elements interacting via tunable Coulomb coupling. We derive the full quantum Langevin equations, perform a linearization procedure to examine fluctuation dynamics, and analyze both complete and phase synchronization between the two subsystems. Our results demonstrate that strong Coulomb interaction can induce robust synchronization even in the absence of direct optical coupling between the cavities. This highlights the potential of electrostatic interactions as a versatile tool for controlling quantum synchronous behavior. By bridging optomechanics and electrostatic coupling, our study offers new insights into the design and control of synchronized quantum networks and provides a foundation for future experiments aiming to harness multi-modal interactions for quantum technologies.

The remainder of this manuscript is structured as follows. \Cref{sec:model} introduces the theoretical model and outlines the derivation of the dynamical equations. \Cref{sec:results} delves into the numerical results. Finally, \Cref{sec:concl} provides concluding remarks.

\section{Theoretical Model and dynamical equations} \label{sec:model}
\begin{figure}[htp!]
	\centering
	\includegraphics[width=1\linewidth]{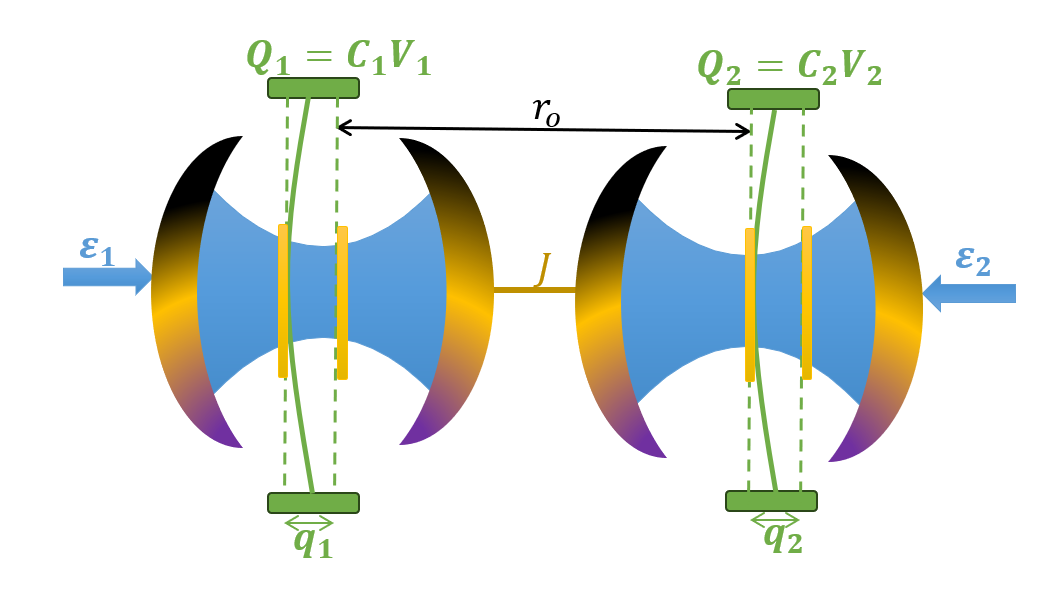}
	\caption{A schematic diagram of two cavities coupled via radiation pressure with strength J and driven by the coupling fields of amplitudes $\mathcal{E}_1$ and $\mathcal{E}_2$ respectively. Mechanical resonators situated inside the cavities are coupled via Coulomb interaction.}
	\label{fig:setup}
\end{figure}

Our benchmark system consists of a hybrid optomechanical setup made of two  coupled optical cavities via photon tunneling, each of them hosting a mechanical oscillator, which are coupled through Coulomb interacting as depicted in Fig. \ref{fig:setup}.
The total Hamiltonian of the system under rotating wave approximation is presented as follows ($\hbar=1$):
\begin{equation}
\begin{aligned}
\mathcal{H}=&\sum_{j=1}^{2}\Delta_{j}a^\dagger_{j}a_{j}+\frac{\omega_{j}}{2}\sum_{j=1}^{2}(q_j^2+p_j^2)-\sum_{j=1,2}g_{j}a^\dagger_{j}a_{j}q_j+\mathcal{H}_{c}+\\&J(a_{1}^\dagger a_{2}+a_{1}a_{2}^\dagger)+i\mathcal{E}_{1}(a^\dagger_{1} -a_{1} )+i\mathcal{E}_{2}(a^\dagger_{2}-a_{2}),
\end{aligned}
\end{equation}
where $\Delta_{j}=\omega_{oj}-\omega_l$ is the cavities detuning, $a_j(a_j^\dagger)$ is the annihilation (creation) operators of the cavity modes which satisfy the commutation relation $[a_j,a_j^\dagger]=\delta_{ij}, i,j=1,2$ and $q_j$ and $p_j$ represent the dimensionless position and momentum operators of the jth mechanical resonators with frequency $\omega_{j}$ which satisfy the commutation relation $[q_j,p_j]=i$.  The strength of the laser driving fields applied to the two optical cavities is denoted by $\mathcal{E}_j$, which determines the rate at which photons are injected into each cavity. The optomechanical coupling of each cavity, represented by $g_j$, quantifies the interaction between the cavity field and the motion of its corresponding mechanical resonator, mediating energy exchange between light and mechanics. The two cavities are also directly coupled to each other with a strength $J$, allowing photons to tunnel between them and enabling indirect correlations between the mechanical resonators. In addition, the term $\mathcal{H}_c$ describes the interaction energy between the two charged mechanical resonators, which arises from the Coulomb force and provides a direct mechanical coupling channel that can influence the collective dynamics of the system. Together, these parameters govern the interplay of optical, mechanical, and electrostatic interactions, forming the basis for the hybrid optomechanical dynamics studied in this work. The charges on two mechanical resonators are respectively $C_1V_1$ and $C_2V_2$, with $C_1(C_2)$ and $V_1(V_2)$ representing the capacitance and voltage of the bias gate respectively. The expression for Coulomb interaction takes the form 
\begin{equation}
\label{Eq2}
\begin{aligned}
\mathcal{H}_{c}&=\frac{C_{1}V_{1}C_{2}V_{2}}{4\pi\epsilon_{o}|r_{o}+q_{1}-q_{2}|}\\&=\frac{C_{1}V_{1}C_{2}V_{2}}{4\pi\epsilon_{o}r_{o}}\left[1-\frac{q_{1}-q_{2}}{r_{o}}+\left(\frac{q_{1}-q_{2}}{r_{o}}\right)^{2}\right],
\end{aligned}
\end{equation}	
here, the linear term may be cancelled by redefining the equilibrium positions. The quadratic term contains a renormalization of the mechanical frequencies, which is just a small frequency shift of the original frequencies and can be neglected. By further omitting the constant term, we then
obtain the coupling of the mechanical oscillators	
$
\mathcal{H}_{c}=\chi_{c}q_{1}q_{2},
$
where $\chi_{c}=\frac{C_{1}V_{1}C_{2}V_{2}}{2\pi\epsilon_{o} r_{o}^{3}}$ is the Coulomb coupling between the two oscillators. This description is analogous to the information presented in~\cite{Bai:24,PhysRevA.90.043825,PhysRevA.91.022326}.
\subsection{Dynamical Equations} \label{sec:IIB}
Now, we proceed to derive  the equations for the dynamics of the system under consideration. By employing the Heisenberg equations of motion and taking into account the effects of noise and damping, we obtain the following set of nonlinear quantum Langevin equations for the system under consideration:
\begin{equation}
\label{Eq6}
\begin{aligned}
\dot{q}_j=&\omega_{j}p_j,\\
\dot{p}_j=&-\omega_{j}q_j+g_ja_j^\dagger a_j-\chi_{c}q_{3-j}-\gamma_{mj}p_j+\xi_j\\
\dot{a}_j=&-(i\Delta_{j}+\kappa_j)a_j+ig_ja_jq_j-iJ_ja_{3-j}+E_j+\sqrt{2\kappa_j}a_j^{in},
\end{aligned}
\end{equation}
where $\gamma_{mj}$ is the damping rate of the mechanical oscillators, $\kappa_j$ is the decay rates of the cavities, $a_j^{in}$ is the the vacuum input noise operator with the nonzero correlation function~\cite{walls2008input}. This optical noise operators satisfy the correlation relation,
$
\langle a_j^{\text{in}\dagger}(t)a_j^{\text{in}}(t^\prime)+a_j^{\text{in}}(t^\prime)a_j^{\text{in}\dagger}(t)\rangle=\delta(t-t^\prime)
$, 
and $\xi_j$ represent the Langevin force, which accounts for the effects of Brownian noise. The nonzero correlation function is defined as~\cite{PhysRevA.63.023812}
\begin{equation}
\langle\xi_j(t)\xi_j(t^\prime)\rangle=\frac{\gamma_{mj}}{\omega_{j}}\int\frac{\omega d\omega}{2\pi}e^{-i\omega(t-t^\prime)}\left[\coth\left(\frac{\omega}{2K_BT}\right)+1\right],
\end{equation}
here, $K_B$ represent the Boltzmann constant and T is the environmental temperature. However, quantum effects become evident only when we consider mechanical oscillators with a large quality factor, i.e $Q\gg1$. Based on this, we have the following Markovian approximation on the Brownian noise operators 
\begin{equation}
\langle\xi_j(t)\xi_j(t^\prime)+\xi_j(t^\prime)\xi_j(t)\rangle/2\approx\gamma_{mj}(2n_{thj}+1)\delta(t-t^\prime),
\end{equation}
where the thermal phonon number is define as\\ $n_{thj}=\left[exp{\left(\frac{\hbar\omega_{j}}{K_BT}\right)}-1\right]^{-1}$. Throughout this work, we will assume that $n_{th1}=n_{th2}=n_{th}$.

In what follows, the nonlinear quantum Langevin equations can be linearized by expressing each Heisenberg operator as the sum of its steady-state mean value and an additional fluctuation operator as:
\begin{equation}
\begin{aligned}
q_j=q_{js}+\delta q_j,~~~
p_j=p_{js}+\delta p_j,~~~
a_j=\alpha_j+\delta a_j.
\end{aligned}
\end{equation}
By substituting these equations into the quantum Langevin equations above Eq.~\eqref{Eq6}, we can derive a set of nonlinear algebra equations for the steady-state values, along with a set of quantum Langevin equations for the fluctuation operators. The classical dynamical equations take the form:
\begin{equation}
\label{Eq3}
\begin{aligned}
\dot{q}_{js}=&\omega_{j}p_{js}\\
\dot{p}_{js}=&-\omega_{j}q_{js}+g_{j}|\alpha_j|^2-\chi_{c}q_{(3-j)s}-\gamma_{mj}p_{js},\\
\dot{\alpha}_j=&-(i\Delta^\prime_{j}+\kappa_j)\alpha_j-iJ\alpha_{3-j}+\mathcal{E}_j,
\end{aligned}
\end{equation}
where $\Delta_j^\prime=\Delta_j-g_jq_{js}$ is the effective optomechanical detuning. The linearized quantum Langevin equations that characterized the fluctuations are presented as follows:
\begin{equation}
\label{Eq9}
\begin{aligned}
\delta\dot{q}_j=&\omega_{j}\delta p_j,\\
\delta\dot{p}_j=&-\omega_{j}\delta q_j+\text{Re}[G_j](\delta a_j+\delta a_j^\dagger)-i\text{Im}[G_j](\delta a_j-\delta a_j^\dagger)\\&-\chi_{c}\delta q_{3-j}-\gamma_j\delta p_j+\xi_j,\\
\delta\dot{a}_j=&-(i\Delta^\prime_{j}+\kappa_j)\delta a_j+i\text{Re}[G_j]\delta q_j-\text{Im}[G_j]\delta q_j-iJ\delta a_{3-j}\\&+\sqrt{2\kappa_j}a_j^{in}\\
\end{aligned}
\end{equation}
where $G_j=g_j\alpha_j$ is the effective optomechanical coupling.

To explore quantum synchronization between the two nanomechanical resonators, we aim to examine the behavior of the figure of merit $S_c $ given by the expression ~\cite{PhysRevLett.111.103605} 
\begin{equation}
S_c(t)=\langle q_{-}(t)^2+p_{-}(t)^2\rangle^{-1}
\end{equation}
where
$
q_{-}(t)=[q_1(t)-q_2(t)]/\sqrt{2},$ and  $p_{-}(t)=[p_1(t)-p_2(t)]/\sqrt{2}.
$
The measure of quantum phase synchronization can be gotten through 
\begin{equation}
S_p(t)=\frac{1}{2}{\langle(\delta p^\prime_{-}(t))^2\rangle}^{-1}
\end{equation}
with 
\begin{equation}
p^\prime_{-}(t)=[p^\prime_1(t)-p^\prime_2(t)]/\sqrt{2},
\end{equation}
and
\begin{equation}
q_j^\prime=q_j\cos\varphi_j+p_j\sin\varphi_j,~~p_j^\prime=p_j\cos\varphi_j-q_j\sin\varphi_j,
\end{equation}
here, the phase $\varphi_j=\arctan\left(\frac{ p_{js}}{ q_{js}}\right)$

At this juncture, it is necessary to define the position and momentum quadrature basis $\delta x_j=\frac{\delta a_j+\delta a_j^\dagger}{\sqrt{2}}$ and $\delta y_j=\frac{\delta a_j-\delta a_j^\dagger}{i\sqrt{2}}$. We can now calculate $S_c$ and $S_p$ by using the covariance matrix of fluctuations. The set of Eq.~\eqref{Eq9}, describing these fluctuation can be written in a compact form as 
\begin{equation}
\label{Eq4}
\dot{z}=A(t)z(t)+\eta(t).
\end{equation}
where A is the drift matrix defined by
\begin{widetext} 
\begin{equation}
A=
\begin{pmatrix}
0&\omega_{1}&0&0&0&0&0&0\\
-\omega_{1}&-\gamma_1&\sqrt{2}\text{Re}[G_1]&\sqrt{2}\text{Im}[G_1]&-\chi_{c}&0&0&0\\
-\sqrt{2}\text{Im}[G_1]&0&-\kappa_1&\Delta_{1}^\prime&0&0&0&J\\
\sqrt{2}\text{Re}[G_1]&0&-\Delta_{1}^\prime&-\kappa_1&0&0&-J&0\\
0&0&0&0&0&\omega_{2}&0&0\\
-\chi_{c}&0&0&0&-\omega_{2}&-\gamma_2&\sqrt{2}\text{Re}[G_2]&\sqrt{2}\text{Im}[G_2]\\
0&0&0&J&-\sqrt{2}\text{Im}[G_2]&0&-\kappa_2&\Delta_{2}^\prime\\
0&0&-J&0&\sqrt{2}\text{Re}[G_2]&0&-\Delta_{2}^\prime&-\kappa_2\\
\end{pmatrix}.
\end{equation}
\end{widetext}
A formal solution of Eq.~\eqref{Eq4} is given by,
\begin{equation}
z(t)=\mathcal{M}(t)z(0)+\int_{0}^{t}dt^{\prime}\mathcal{M}(t^{\prime})\mathcal{N}(t-t^{\prime}),
\end{equation}
where $\mathcal{M}(t)=\exp(At)$.
Here, the quadrature fluctuation vector is given by $z^\top(t)=(\delta x_{1},\delta y_{1},\delta x_{2},\delta y_{2},\delta q_1, \delta p_1,\delta q_2,\delta p_2)$ and $\eta^\top=(\sqrt{2\kappa_1}x_1^{\text{in}},\sqrt{2\kappa_1}y_1^{\text{in}},\sqrt{2\kappa_2}x_2^{\text{in}}, \sqrt{2\kappa_2}y_2^{\text{in}},0,\xi_1,0,\xi_2)$ is the noise vector. The input noise operators of the cavity fields are expressed in terms of quadratures as
\begin{equation}
x^{\text{in}}_j = \frac{a^{\text{in}}_j + a_j^{\text{in}\dagger}}{\sqrt{2}}, \qquad
y^{\text{in}}_j = \frac{a^{\text{in}}_j - a_j^{\text{in}\dagger}}{i\sqrt{2}},
\end{equation}
where $x^{\text{in}}_j$ and $y^{\text{in}}_j$ represent the amplitude and phase quadrature components of the input noise for the $j$th cavity.
The following covariance matrix (CM) can enable us to explore the contribution of quantum fluctuation to quantum complete synchronization, $\phi$-synchronization and phase synchronization of the system under consideration:
\begin{equation}
V_{ij}=\left[\langle z_i(t)z_j(t^\prime)+z_j(t^\prime)z_i(t)\rangle\right]/2.
\end{equation}
The evolution of CM over time is govern by 
\begin{equation}
\label{Eq5}
\dot{V}(t)=A(t)V(t)+V(t)A(t)^\top+\mathcal{D}
\end{equation}
where $\mathcal{D}$ is the diffusion matrix define by 
\begin{equation}
\mathcal{D}=\text{diag}(0,\gamma_{m1}(2n_{th1}+1),\kappa_1,\kappa_1,0,\gamma_{m2}(2n_{th2}+1),\kappa_2,\kappa_2).
\end{equation} 
At this juncture, we can write the concise form of complete quantum synchronization, $\phi$-synchronization synchronization and quantum phase synchronization as 
\begin{equation}
\begin{aligned}
S_c=&[1/2(V_{11}(t)+V_{22}(t)+V_{55}(t)+V_{66}(t)-2V_{15}(t)-2V_{26}(t))]^{-1},\\
S^\phi_c=&[1/2(V_{11}(t)+V_{22}(t)+V_{55}(t)+V_{66}(t)+2V_{25}(t)\sin\phi\\&-2V_{16}(t)\sin\phi-2V_{26}\cos\phi-2V_{15}(t)\cos\phi)]^{-1}\\
S_p=&[(V_{11}(t)\sin^2\phi_1+ V_{22}(t)\cos^2\phi_1+V_{55}(t)\sin^2\phi_2+\\&V_{66}(t)\cos^2\phi_2-2V_{12}(t)\cos\phi_1\sin\phi_1-2V_{15}(t)\sin\phi_1\sin\phi_2+\\&2V_{16}\sin\phi_1\cos\phi_2+2V_{25}(t)\cos\phi_1\sin\phi_2-\\&2V_{26}(t)\cos\phi_1\cos\phi_2-2V_{56}(t)\cos\phi_2\sin\phi_2)]^{-1}.
\end{aligned}
\end{equation}
\section{Numerical results}\label{sec:results}
In this section, we investigate the dynamics of the system under consideration by examining the time evolution of the mean values of the position and momentum operators associated with the two mechanical resonators and cavities. This analysis allows us to gain insight into the coupled motion of the resonators and the influence of the system parameters on their dynamical behaviour. We further study different synchronization regimes, including complete quantum synchronization, $\phi$-synchronization, and quantum phase synchronization.

 \subsection{Classical dynamics} \label{sec:IIIA}
To analyze the dynamics of the mechanical resonators, we solve a set of Eq.~\eqref{Eq3} numerically by setting initial conditions to zero i.e the initial state of the cavities is vacuum. It is worth mentioning that the mechanical phase of each resonator can be defined from the mean values as shown below 
\begin{equation}
\varphi_j=\arctan\left(\frac{ p_{js}}{ q_{js}}\right)
\end{equation}
This represents the angular position of the mechanical trajectory in phase-space. The classical synchronization is identified by the locking of the phase difference, such that $\Delta\varphi=\varphi_1-\varphi_2$ remains constant in time.

\begin{figure}[htp!]
	\centering
	{\includegraphics[width=4.2cm]{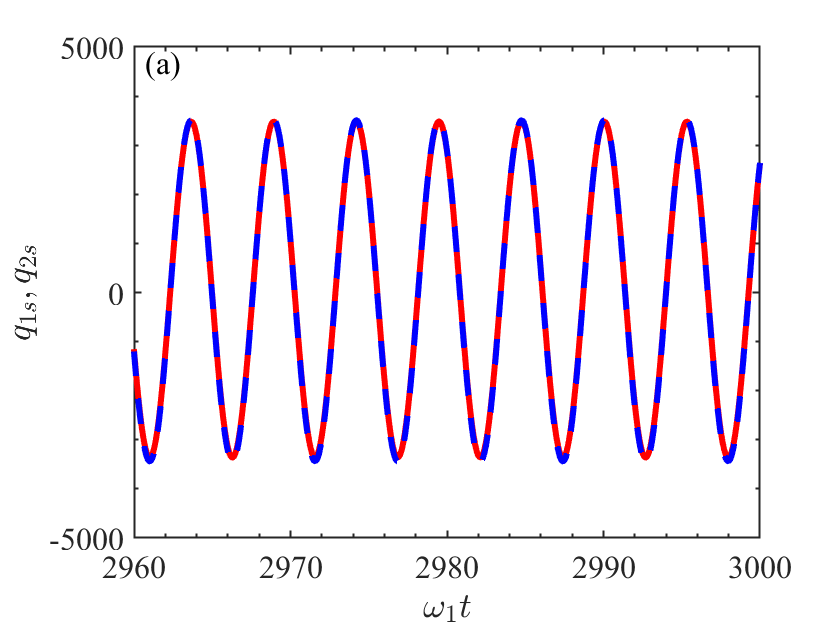}}
	{\includegraphics[width=4.2cm]{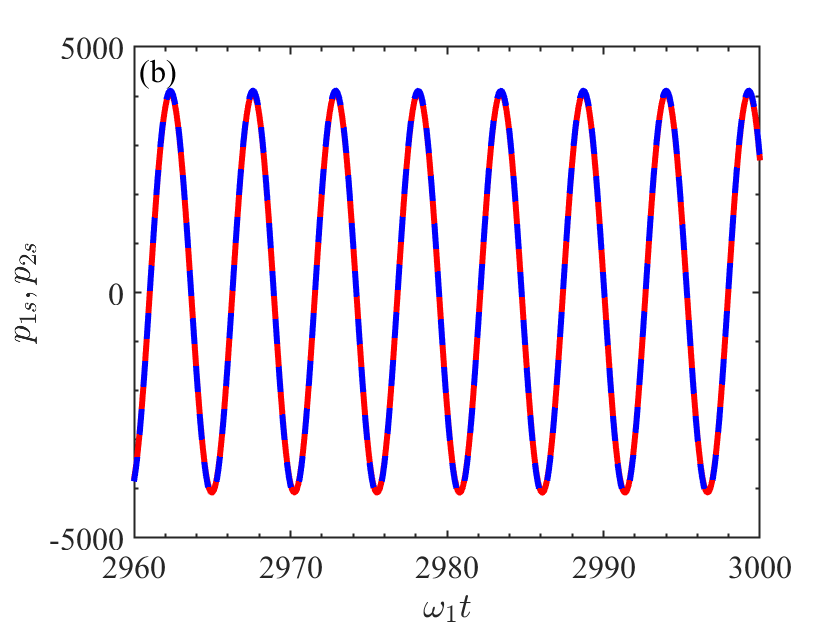}}
	{\includegraphics[width=4.2cm]{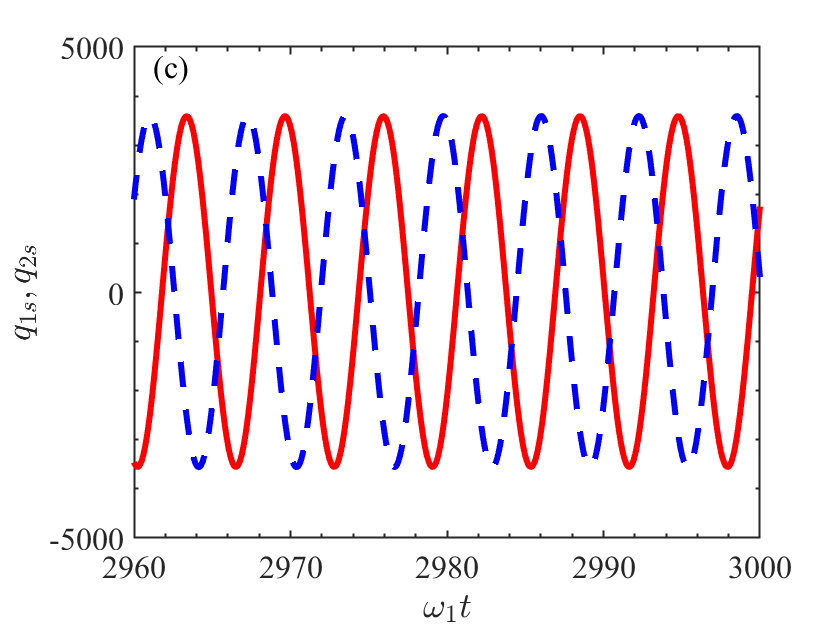}}
	{\includegraphics[width=4.2cm]{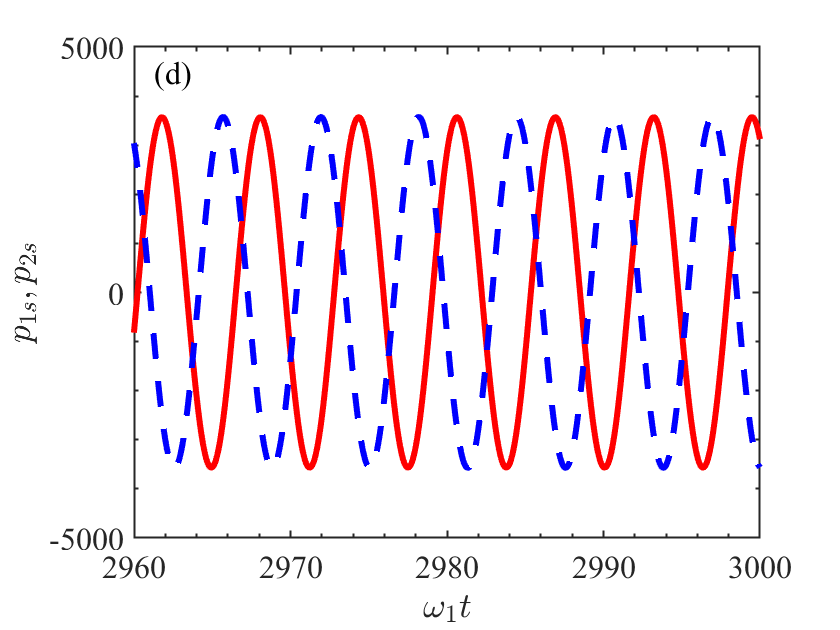}}
	\caption{Variation of the mean values of the mechanical resonators for different Coulomb coupling strengths. (a) $q_{1s}$ (red solid line) and $q_{2s}$ (blue dashed line) for $\chi_c/\omega_{1}=0.4$, (b) $p_{1s}$ (red solid line) and $p_{2s}$ (blue dashed line) for $\chi_c/\omega_{1}=0.4$, (c) $q_{1s}$ (red solid line) and $q_{2s}$ (blue dashed line) for $\chi_c/\omega_{1}=0.0$, and (d) $p_{1s}$ (red solid line) and $p_{2s}$ (blue dashed line) for $\chi_c/\omega_{1}=0.0$. Parameters are chosen as: $\omega_2/\omega_1=1.005$, $\Delta_1=-\omega_1$, $\Delta_2=-\omega_2$,
		$g_1/\omega_1=1\times 10^{-3}$, $g_2=g_1$, $\gamma_{mj}/\omega_{1}=1\times 10^{-3}$, $\kappa_1/\omega_1=0.15$, $\kappa_2=\kappa_1$, $J/\omega_{1}=0.02$,  $\mathcal{E}_1/\omega_{1}=150$, $\mathcal{E}_2=\mathcal{E}_1$.}
	\label{fig:2}	
\end{figure}

 As illustrated in Fig.~\ref{fig:2}(a--b), the mean values of the position operators 
$q_{1s}(t)$ and $q_{2s}(t)$, together with the corresponding momentum mean values $p_{1s}(t)$ and $p_{2s}(t)$, exhibit perfectly synchronized oscillations, remaining exactly in phase throughout the stable dynamical regime. This in-phase behaviour indicates strong coherent coupling between the two mechanical resonators. By contrast, as shown in Fig.~\ref{fig:2}(c--d), when the Coulomb coupling between the two mechanical resonators is absent, their dynamical behaviour changes significantly. The resonators oscillate out of phase, demonstrating a loss of synchronization induced by the removal of the inter-resonator interaction. This clearly shows that Coulomb interaction is the key ingredient for synchronization in our system. In other words, strong Coulomb coupling between the mechanical resonators results in a coherent exchange of energy between the mechanical resonators for a long time and therefore the oscillations do not decay down but constant for a long time.

\begin{figure}[htp!]
	\centering
	{\includegraphics[width=4.2cm]{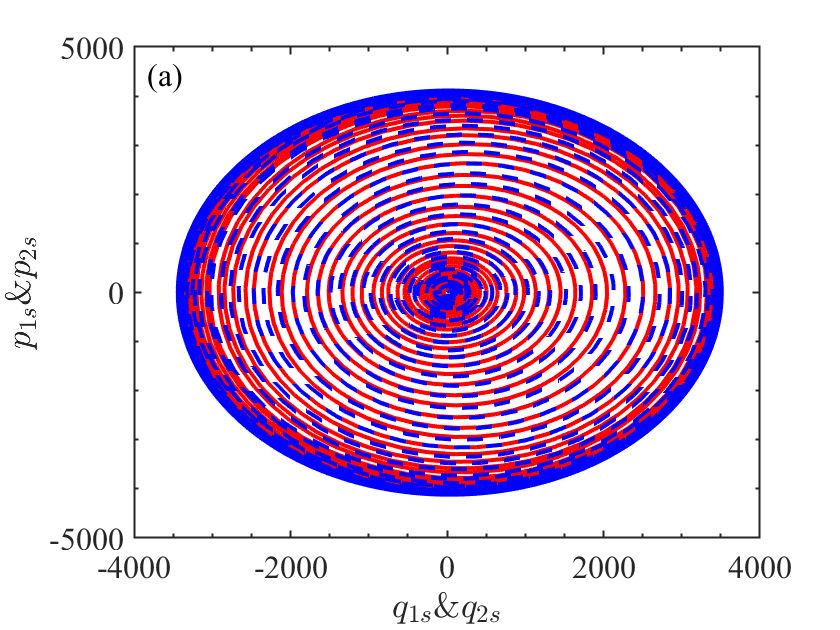}}
	{\includegraphics[width=4.2cm]{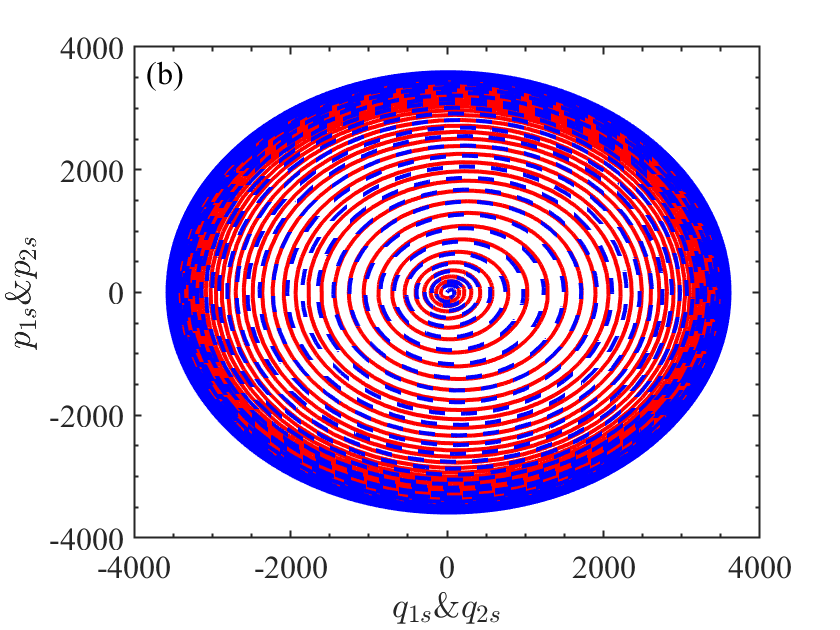}}
	\caption{ (a) The limit cycle trajectories in $ q_{1s}(t)\rightleftharpoons p_{1s}(t)$ and $ q_{2s}(t)\rightleftharpoons p_{2s}(t)$ spaces for $\chi_c/\omega_{1}=0.4$. (b) The limit cycle trajectories in $ q_{1s}(t)\rightleftharpoons p_{1s}(t)$ and $ q_{2s}(t)\rightleftharpoons p_{2s}(t)$ spaces for $\chi_c/\omega_{1}=0.0$. The parameters are as in Fig.~\ref{fig:2}.}
	\label{fig:3}	
\end{figure}

As illustrated in Fig.~\ref{fig:3}, the phase-space trajectories defined by the pairs 
$q_{1s}(t)\leftrightharpoons p_{1s}(t)$ and $q_{2s}(t)\leftrightharpoons p_{2s}(t)$  for the two nanomechanical resonators evolve toward stable asymptotic periodic orbits. This behaviour indicates that, after a transient dynamical regime, each resonator settles into a limit-cycle motion characterized by sustained and regular oscillations.

\begin{figure}[htp!]
	\setlength{\lineskip}{0pt}
	\centering
	{\includegraphics[width=4.2cm]{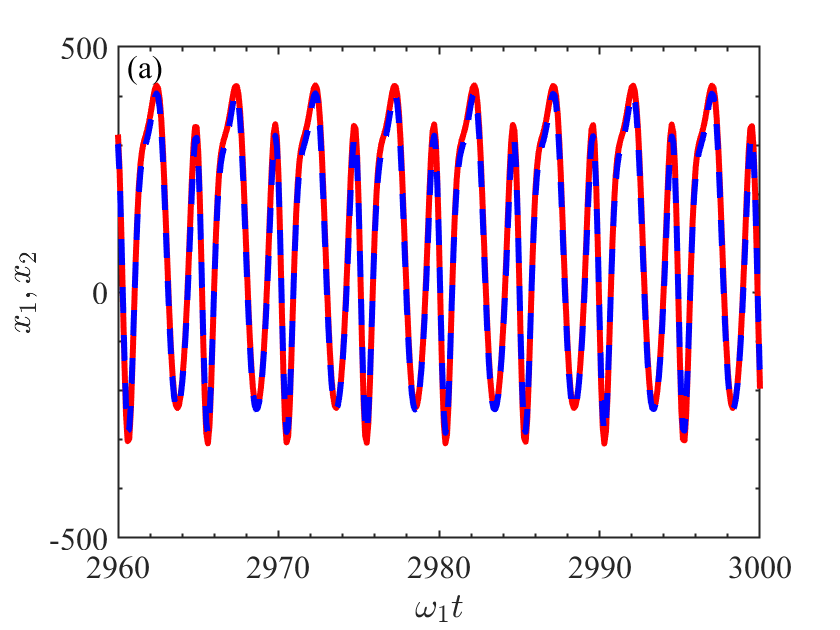}}
	{\includegraphics[width=4.2cm]{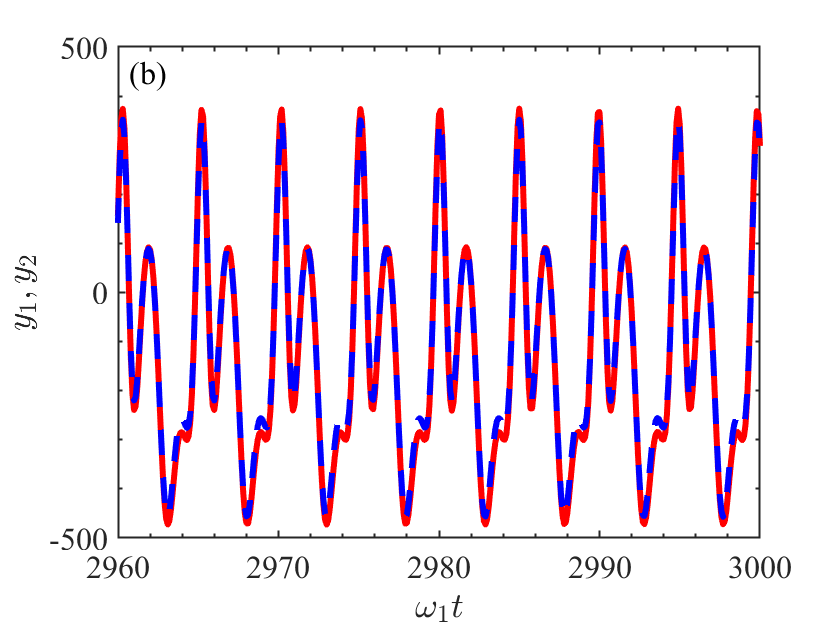}}
	\caption{ (a) The time evolution of position and momentum of the two cavities (a) $x_{1}$ (red solid line), $ x_{2}$ (blue dashed line). (b) $ y_{1}$ (red solid line), $ y_{2}$ (blue dashed line). The parameters are as in Fig.~ \ref{fig:2} except for $J/\omega_1=0$ and $\chi_c/\omega_{1}=0.6$.}
	\label{fig:4}	
\end{figure}

In Fig.~\ref{fig:4}, we show the time evolution of position and momentum operators of the two cavity fields in the absent of the coupling constant $J$. We found that the two cavities oscillates in phase in a long time when they are not directly coupled but communicating via the two mechanical resonators in the presence of strong coulomb interaction. This shows that their oscillations synchronize. This enable significant energy exchange between the two cavity fields since their crests and troughs align over a long time. As can be seen from Fig.~\ref{fig:4}, this energy exchange fluctuate at some points since the amplitude of oscillation varies. An analogous behaviour is observe with the mechanical resonators (see Fig.~\ref{fig:2}(a-b)) but the amplitude is constant through out the oscillation. This shows that the rate at which energy is transferred between the resonators is constant. The comparison highlights the complementary roles of the two subsystems: the mechanical resonators act as stable mediators of interaction, while the cavity fields, being indirectly coupled (see Fig.~\ref{fig:4}(a-b)), exhibit amplitude-dependent fluctuations in their energy exchange process.

\begin{figure}[htp!]
	\setlength{\lineskip}{0pt}
	\centering
	{\includegraphics[width=4.2cm]{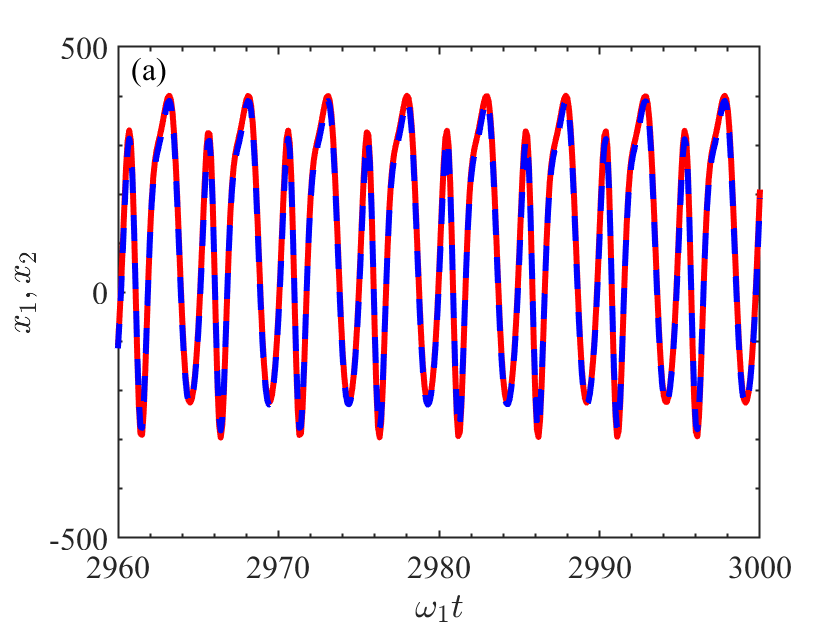}}
	{\includegraphics[width=4.2cm]{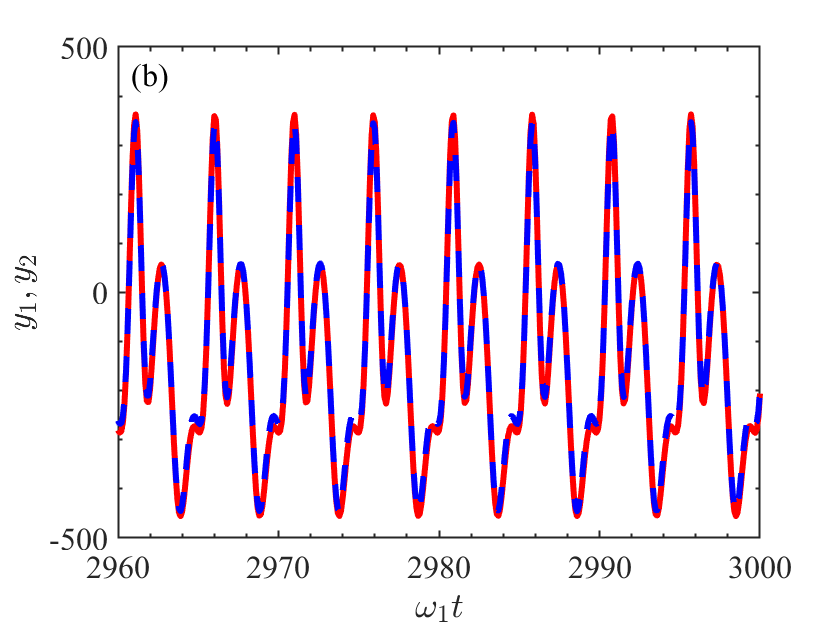}}
	\caption{ (a) The time evolution of position and momentum of the two cavities (a) $x_{1}$ (red solid line), $ x_{2}$ (blue dashed line). (b) $ y_{1}$ (red solid line), $ y_{2}$ (blue dashed line). The parameters are as in Figu.~ \ref{fig:2} except for $J/\omega_1=0.02$ and $\chi_c/\omega_{1}=0.6$.}
	\label{fig:5}	
\end{figure}

 As illustrated in Fig.~\ref{fig:5} we now considered a small direct coupling between the two cavities ($J/\omega_1= 0.02$). The overall behaviour remains similar to the $J/\omega_1=0$ case: the cavities oscillate in phase and efficiently exchange energy with fluctuating amplitude. This is because synchronization is primarily mediated by the mechanical resonators via strong Coulomb interaction, and the weak direct coupling only slightly reinforces the phase coherence without significantly altering the dynamics.

\begin{figure}[htp!]
	\setlength{\lineskip}{0pt}
	\centering
	{\includegraphics[width=4.2cm]{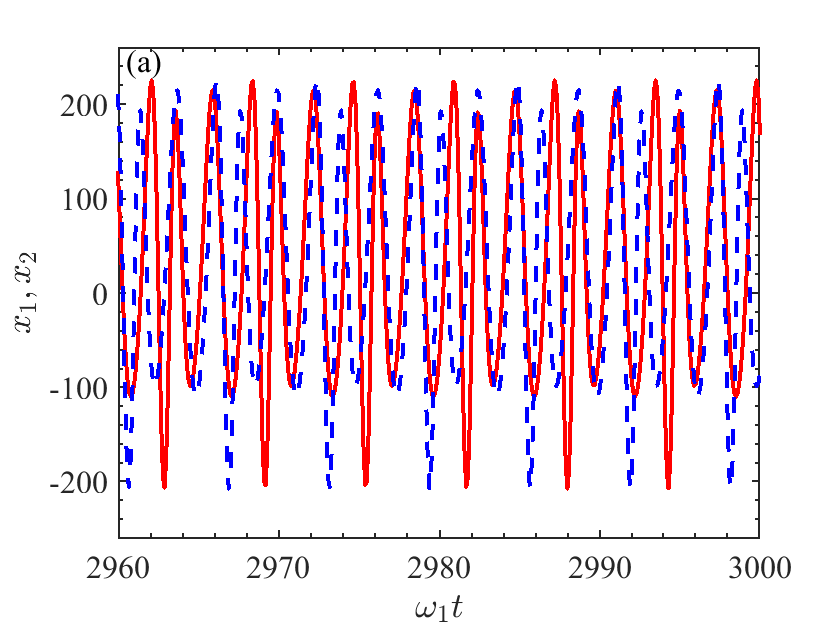}}
	{\includegraphics[width=4.2cm]{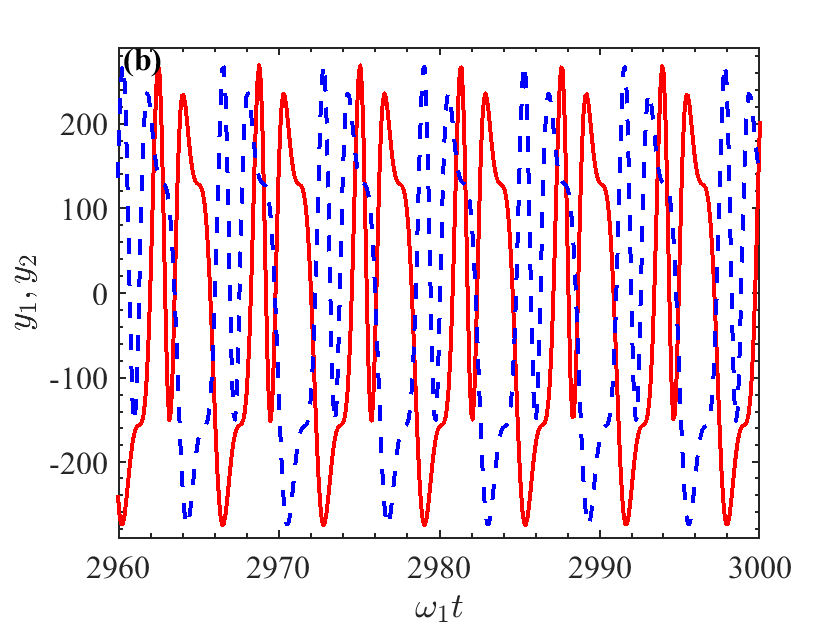}}
	\caption{ (a) The time evolution of position and momentum of the two cavities (a) $x_{1}$ (red solid line), $ x_{2}$ (blue dashed line). (b) $ y_{1}$ (red solid line), $ y_{2}$ (blue dashed line). The parameters are as in Fig.~ \ref{fig:2} except for $J/\omega_1=0.02$ and $\chi_c/\omega_{1}=0.0$.}
	\label{fig:6}	
\end{figure}
As can be seen from Fig.~\ref{fig:6}, in the absence of Coulomb interaction, the cavities oscillate out of phase despite being directly coupled. This behaviour can be explained by the fact that, without Coulomb interaction, the mechanical resonators are unable to mediate in-phase synchronization between the cavities.
Based on the above observations, both the cavity fields and the mechanical resonators exhibit classical synchronization in the presence of Coulomb interaction. The Coulomb-mediated coupling acts as a crucial bridge, ensuring long-term phase coherence and coordinated dynamics across the entire hybrid optomechanical configuration. This synchronization is essential for efficient energy transfer, as phase alignment optimizes the exchange of energy between the mechanical and optical degrees of freedom. It is also vital for coherent control: by stabilizing the relative phases of the resonators, the Coulomb interaction provides an effective tuning knob for manipulating the collective state of the system. Moreover, the emergence of classical-like synchronization in the system’s first moments establishes a stable dynamical background that facilitates the preservation of quantum synchronization.

\subsection{Complete quantum synchronization, $\phi$-synchronization and quantum phase synchronization} \label{sec:IIIB}
In this section, we commence the numerical analysis with the solution of Eq.~\eqref{Eq5} which characterize the behaviour of the CM elements associated with the optical and mechanical modes. For the initial conditions, we assumed that the cavities vacuum state and the mechanical resonators are prepared in the thermal state during the evolution of the system i.e. $V(0)=\text{diag}[1/2,1/2,1/2,1/2,n_{th}+1/2,n_{th}+1/2,n_{th}+1/2,n_{th}+1/2]$.
\begin{figure*}[htp!]
	\setlength{\lineskip}{0pt}
	\centering
	{\includegraphics[width=5.5cm]{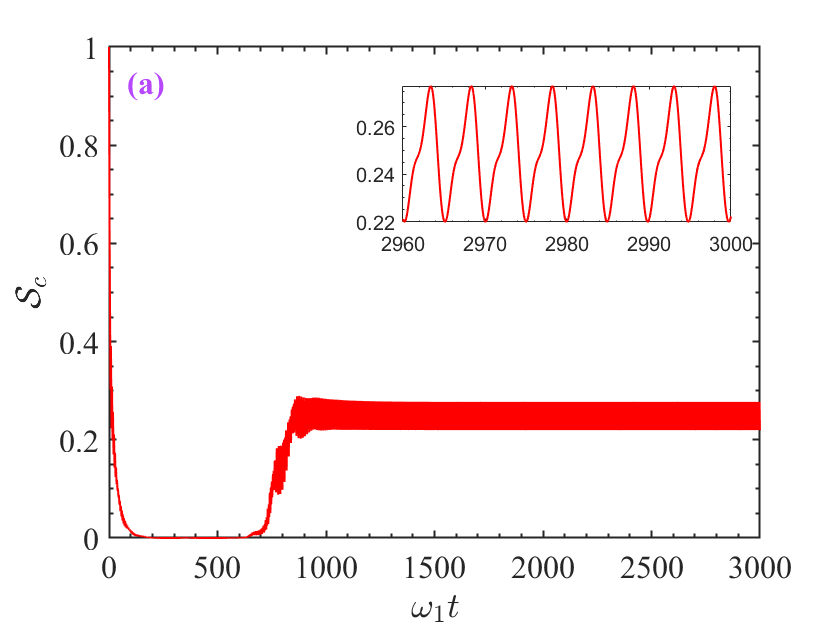}}
	{\includegraphics[width=5.5cm]{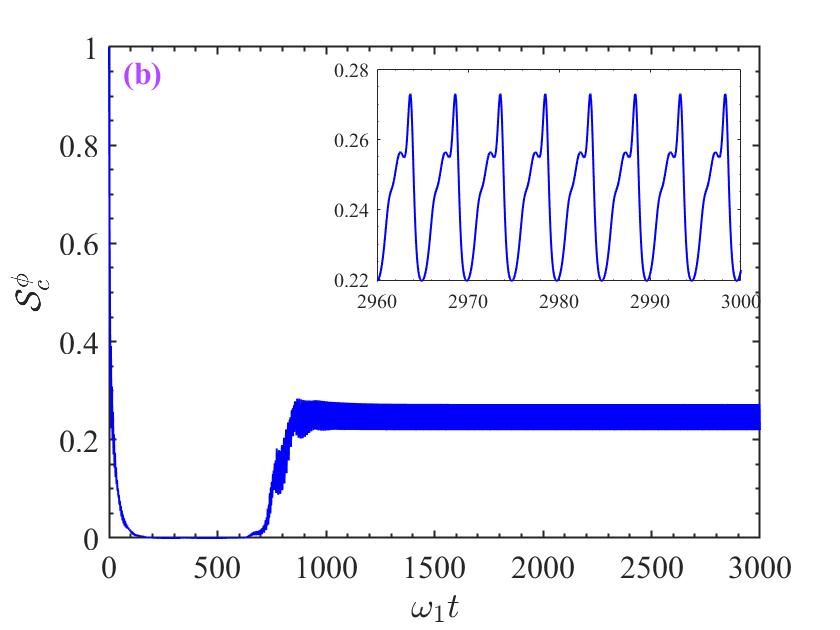}}
	{\includegraphics[width=5.5cm]{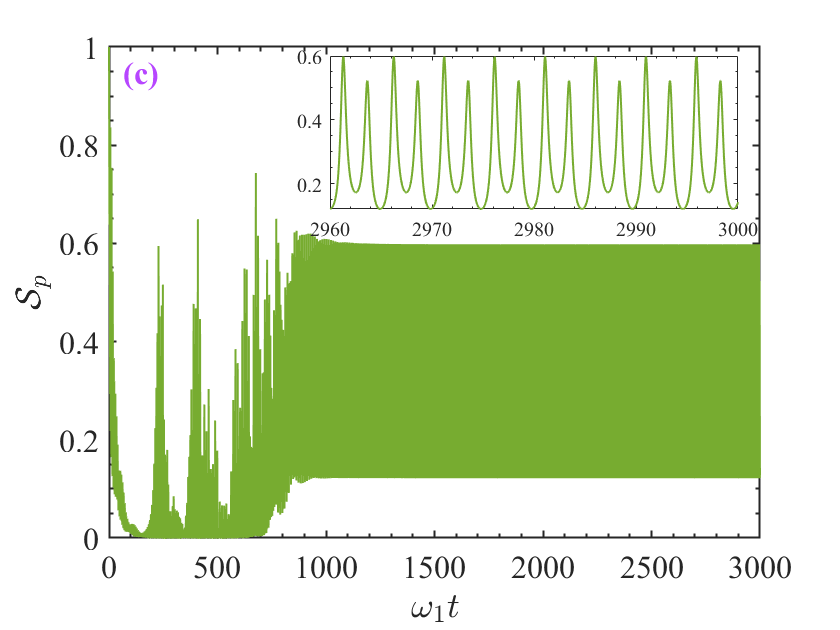}}
	\caption{Time evolution of (a) complete quantum synchronization, (b) $\phi$ synchronization and (c) quantum phase-synchronization. Parameters are chosen as: $\omega_2/\omega_1=1.005$, $\Delta_1=-\omega_1$, $\Delta_2=-\omega_2$,
		$g_1/\omega_1=1\times 10^{-3}$, $g_2=g_1$, $\gamma_{mj}/\omega_{1}=1\times 10^{-3}$, $\kappa_1/\omega_1=0.15$, $\kappa_2=\kappa_1$, $J/\omega_{1}=0.02$,  $\mathcal{E}_1/\omega_{1}=150$, $\mathcal{E}_2=\mathcal{E}_1$, $\chi_c/\omega_{1}=0.6$, and $n_{th}=0$.}
	\label{fig:7}	
\end{figure*}
In Fig.~\ref{fig:7}, we provide a comparative illustration of complete quantum synchronization, $\phi$-synchronization, and quantum phase synchronization as they evolve between two mechanical resonators. Looking at this figure carefully, one can clearly see that the system start with an initial transient and later settle down to a steady state. The dynamics of quantum complete synchronization and $\phi$-synchronization differ notably from those of quantum phase synchronization, as illustrated in Fig.~\ref{fig:7}. While complete and $\phi$-synchronization involve correlations in both amplitude and phase of the mechanical resonators, quantum phase synchronization focuses primarily on the alignment of their phases, regardless of amplitude correlations.

\begin{figure*}[htp!]
	\setlength{\lineskip}{0pt}
	\centering
	{\includegraphics[width=5.5cm]{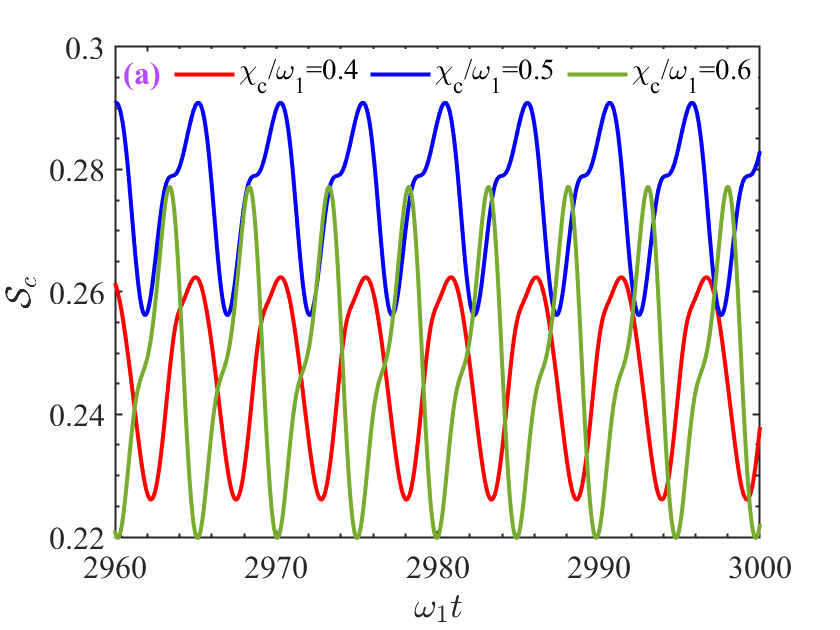}}
	{\includegraphics[width=5.5cm]{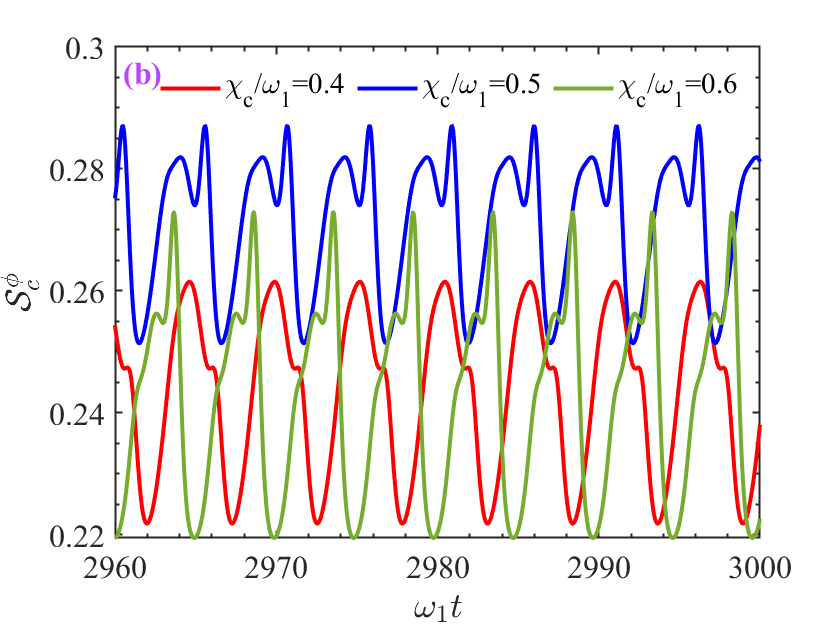}}
	{\includegraphics[width=5.5cm]{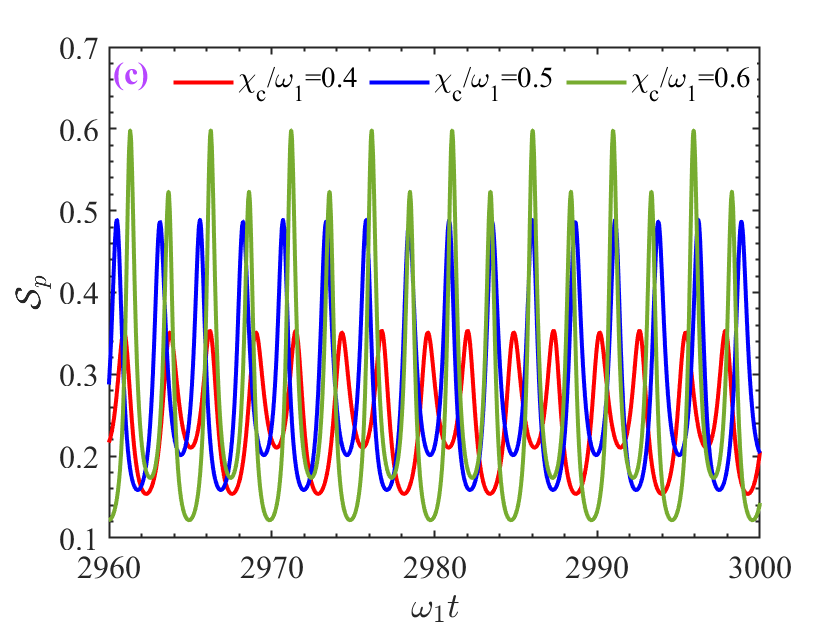}}
	\caption{Time evolution of (a) complete quantum synchronization, (b) $\phi$ synchronization and (c) quantum phase-synchronization for different values of the Coulomb coupling $\chi_c$. The figure highlights how different types of synchronization behave under varying system parameters. Parameters are chosen as: $\omega_2/\omega_1=1.005$, $\Delta_1=-\omega_1$, $\Delta_2=-\omega_2$,
		$g_1/\omega_1=1\times 10^{-3}$, $g_2=g_1$, $\gamma_{mj}/\omega_{1}=1\times 10^{-3}$, $\kappa_1/\omega_1=0.15$, $\kappa_2=\kappa_1$, $J/\omega_{1}=0.02$,  $\mathcal{E}_1/\omega_{1}=150$, $\mathcal{E}_2=\mathcal{E}_1$, and $n_{th}=0$.}
	\label{fig:8}	
\end{figure*}
It shows clearly from Fig.~\ref{fig:8} that, quantum phase synchronization is enhanced by increasing the Coulomb coupling between the two mechanical resonators. This demonstrates that Coulomb interaction plays a pivotal role in promoting phase coherence between the resonators, effectively aligning their quantum phases over time. The observation highlights the importance of inter-resonator Coulomb interaction as a key mechanism for controlling and optimizing quantum phase synchronization in hybrid optomechanical systems.
\subsection{Experimental feasibility}
Our proposed optomechanical architecture can be realized with existing experimental technology. Coupled-cavity optomechanical systems have been demonstrated in both optical and microwave platforms~\cite{Aspelmeyer2014,Lecocq2015}, where radiation-pressure interaction between cavity photons and localized mechanical modes is routinely achieved. Photon tunneling between cavities can be implemented using evanescent fields, optical fibers, or capacitive links, depending on the physical platform~\cite{Not2006}. More importantly, the feasibility of strong and tunable electrostatic (Coulomb) coupling involving mechanical oscillators has been firmly established. Tunable electrostatic coupling between nano- and electromechanical resonators has been experimentally demonstrated using bias voltages~\cite{ILYAS2016,Zhang}. Additionally, in Ref~\cite{Hen2015} it can be clearly seen that a mesoscopic mechanical cantilever can be strongly and controllably coupled via Coulomb interaction to a trapped charged particle using experimentally accessible bias voltages, achieving coupling strengths in the tens of kilohertz regime and explicitly demonstrating that such electrostatic coupling schemes are feasible with existing technology. The coupling mechanism, tunability, and parameter regime demonstrated in that work directly support the realization of Coulomb-mediated interactions between nearby charged mechanical oscillators in the present proposal~\cite{Vign}. Typical experimental parameters mechanical such as frequencies in the kHz-MHz range, Coulomb coupling strengths of several kHz, and controllable optomechanical coupling rates are well within experimentally demonstrated regimes~\cite{Cast2012}. These considerations collectively indicate that the proposed hybrid optomechanical system is experimentally feasible with current technology.
\section{conclusion}\label{sec:concl}
In summary, we have investigated the impact of Coulomb interaction on complete quantum synchronization, $\phi$-synchronization and quantum phase synchronization within a four-mode optomechanical configuration. Our results reveal a clear distinction between these regimes. While quantum phase synchronization is enhanced by the Coulomb coupling between the mechanical resonators, the interaction has no substantial effect on complete quantum or $\phi$-synchronization. These findings demonstrate that the Coulomb interaction is the primary mechanism for inducing phase coherence in this system, as it facilitates the exchange of energy between the mechanical resonators and the cavity fields. Because complete and $\phi$-synchronization are largely governed by the internal optomechanical driving, the Coulomb force acts specifically as a tuner for phase alignment. Our work provides a framework for controlling collective quantum dynamics, which may find valuable applications in quantum communication and synchronization-based quantum networks.

\section*{Acknowledgments}
P.D. acknowledges the Iso-Lomso Fellowship at Stellenbosch Institute for Advanced Study (STIAS), Wallenberg Research Centre at Stellenbosch University, Stellenbosch 7600, South Africa, and The Institute for Advanced Study, Wissenschaftskolleg zu Berlin, Wallotstrasse 19, 14193 Berlin, Germany. J.-X.P. is supported by National Natural Science Foundation of China (Grant No.~12504566), Natural Science Foundation of Jiangsu Province (Grant No.~BK20250947), Natural Science Foundation of the Jiangsu Higher Education Institutions (Grant No.~25KJB140013) and Natural Science Foundation of Nantong City (Grant No.~JC2024045).  The authors are thankful to the Deanship of Graduate Studies and Scientific Research at University of Bisha for supporting this work through the Fast-Track Research Support Program.

\textbf{Author Contributions:} E.K.B. and P.D. conceptualized the work and carried out the simulations and analysis. J.-X.P and S.K.S participated in all the discussions and provided useful methodology and suggestions for the final version of the manuscript. J.G and A.S. participated in writting the manuscript and the validation.  A.-H.A.-A., and S.G.N.E. participated in the discussions and supervised the work. All authors participated equally in the writing, discussions, and the preparation of the final version of the manuscript.

\textbf{Competing Interests:} All authors declare no competing interests.

\textbf{Data Availability:}
Relevant data are included in the manuscript and supporting information. 
\bibliography{RefnSSyn}
\end{document}